\documentclass[british,3p, review, number, sort&compress, margin=2.54cm, 12pt]{elsarticle}
\usepackage[T1]{fontenc}
\usepackage[latin9]{inputenc}
\usepackage{babel}
\usepackage{amssymb}
\usepackage{graphicx}
\usepackage{setspace}
\usepackage{esint}
\usepackage[unicode=true,
 bookmarks=true,bookmarksnumbered=false,bookmarksopen=false,
 breaklinks=false,pdfborder={0 0 1},backref=false,colorlinks=false]
 {hyperref}

\makeatletter

\providecommand{\tabularnewline}{\\}

\date{}

\makeatother

\begin{document}

\begin{frontmatter}{}

\title{Modeling the Physical Properties of Environmentally-Friendly Optical
Magnetic Switches: DFT and TD-DFT}

\author{Lat\'evi Max LAWSON DAKU}

\address{Universit\'e de Gen\`eve -- Sciences II, Quai Ernest Ansermet
30, 1211 Gen\`eve, SUISSE}

\ead{max.lawson@unige.ch}

\author{Mark Earl CASIDA}

\address{Laboratoire de Spectrom\'etrie, Interactions et Chimie th\'eorique
(SITh), D\'epartement de Chimie Mol\'eculaire (DCM, UMR CNRS/UGA
5250), Institut de Chimie Mol\'eculaire de Grenoble (ICMG, FR2607),
Universit\'e Grenoble Alpes (UGA) 301 rue de la Chimie, BP 53, F-38041
Grenoble Cedex 9, FRANCE}

\ead{mark.casida@univ-grenoble-alpes.fr}

\clearpage{}
\begin{abstract}
The dominant majority of the hundreds of available spin-crossover
compounds, including the technologically most promising ones, are
based on the Earth-abundant metal iron, making these switches particularly
appealing in terms of sustainable technology. Furthermore, it has
recently been established that these materials may be synthesized
using the techniques of Green Chemistry. Spin crossover in transition
metal complexes can be induced by a change of temperature, by the
application of an external pressure, or a magnetic field, and also
by photoexcitation. Given the wide variety of functionalities to which
these bistable photomagnetic systems could give access, they may be
viewed as the prototypes of molecular-scale optomagnetic switches.
Hence, in response to the growing demand for storing and treating
increasingly-dense information, much effort has been devoted over
several decades to the design of transition metal materials exhibiting
spin crossover in technologically-accessible temperature ranges. The
cornerstone for designing new efficient photoactive spin-crossover
materials remains the ability to predict the magnetic behaviour and
the photoresponse of any first-row transition complex, and the manner
in which its properties are influenced by its environment. This is
a challenge for the inorganic chemist and also, most notably, for
the theoretical and the computational chemist. This chapter gives
an overview of the issues tied to the application of DFT and TD-DFT
to the characterisation of transition metal complexes in the framework
of spin-crossover and related phenomena.
\end{abstract}
\begin{keyword}
Transition metal complexes, green chemistry, spin crossover, DFT,
TD-DFT
\end{keyword}

\end{frontmatter}{}

\section{Introduction}

In response to the growing demand for storing increasingly-dense information,
an important effort is being made to develop new materials for information
storage and processing at the molecular level. Transition-metal systems
have a special role in this area because of their extremily diverse
photophysical and photochemical properties and because of the possibility
of coupling these photoproperties to magnetic properties. This is
precisely the situation with spin-crossover (SCO) complexes as they
may be reversibly switched between their low-spin ground state and
their high-spin metastable excited state~--- two states with quite
different optical and magnetic properties. The change of spin state
may be induced by a change in temperature, by applying an external
pressure or a magnetic field, or by photo-excitation. Given the vast
range of functionalities which can in principle be accessed by these
bistable photomagnetic systems, spin-crossover complexes may be considered
to be the archetype of all optical magnetic switches. They have also
been the object of many multidisciplinary studies, which have been
reviewed in \emph{Topics in Current Chemistry} \citep{GG04}. Given
that the dominant majority of the hundreds of available spin-crossover
compounds, including the technologically most promising ones, are
based on the Earth-abundant metal iron, these switches prove to be
particularly appealing in terms of sustainable technology. Furthermore,
mechanochemistry has recently been applied to the synthesis of SCO
materials, thus demonstrating that these materials may be synthesized
using the techniques of Green Chemistry \citep{ChemCommun_54_180}.

The work of the authors on spin-crossover and \textquotedblleft related\textquotedblright{}
complexes has been carried out in the context of the study of photoswitching.
A deeper understanding of this phenomenon requires the ability to
answer the following key questions:
\begin{itemize}
\item $\left(\mathcal{Q}1\right)$ What is the nature of the states involved?
(structures, energies, lifetimes, etc. ...)
\item $\left(\mathcal{Q}2\right)$ How does the system interact with light?
(excitation energies and electromagnetic transition moments)
\item $\left(\mathcal{Q}3\right)$ How does the system evolve after photoexcitation?
\end{itemize}
Obtaining precise answers to these three questions is a major challenge
to which we hope to contribute through the use of density-functional
theory (DFT) and time-dependent DFT (TD-DFT). This chapter gives an
overview of problems related to the application of DFT and of TD-DFT
to the study of spin-crossover complexes using examples taken principally
from the work of the authors and the reader will be referred to the
original papers for more detail.

This chapter is organized as follows: The next section presents a
review of applicable methods with particular emphasis on DFT and on
TD-DFT. In Sec.~\ref{sec:applications:dft}, stock is taken of the
performance of different DFT functionals and it is shown that modern
functionals can compete at a quantitative level with \emph{ab initio}
methods but at less computational cost. In Sec.~\ref{sec:applications:tddft},
the application of TD-DFT to the study of the excited states of complexes
is considered. In Sec.~\ref{sec:environmental-effects}, the important
subject of environmental effects on the behavior of spin-crossover
complexes is tackled. Finally in Sec.~\ref{sec:the-future} the authors
conclude and try to imagine the future evolution of this field of
study.

\section{Methodology: Three Pauling Points\label{sec:methodology}}

\begin{figure}
\begin{centering}
\includegraphics[scale=0.7]{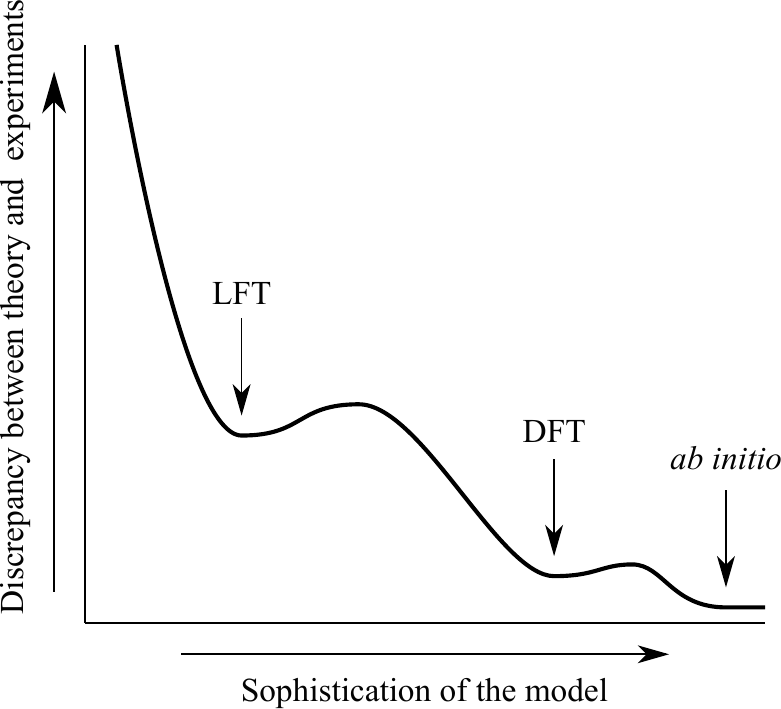}
\par\end{centering}
\caption{\label{fig:Pauling}Evolution of the level of agreement between theory
and experiment as a function of the sophistication of the theoretical
model. The local minima where theory and experiment agree best are
the Pauling points. In the spin-crossover case, we can distinguish
three Pauling points (see Text).}
\end{figure}

``Essentially, all models are wrong, but some are useful \citep{BD87}.''
This quote from American statistician George E.P.\ Box, taken out
of context, reminds us that models are, despite our best efforts,
useful but necessarily imperfect. Very often, during the development
of a model, the level of agreement between theory and experiment does
not increase in a monatomic manner with increasing sophistication
of the model but rather has a tendency to oscillate and local minima
appear (Fig.~\ref{fig:Pauling}.) In the folklore of theoretical
chemistry, these local minima are frequently referred to as Pauling
points because Linus Pauling (Nobel prizes: Chemistry, 1954; Peace,
1962) supposedly had the talent of stopping at these points \citep{pauling}.

Three Pauling points can be identified for spin-crossover complexes.
The first is given by ligand field theory (LFT). This old but well-established
semi-empirical theory is universally used to understand the behaviour
of transition metal complexes \citep{FH00}. It may be regarded as
an essential and widely used language that experimentalists use to
analyse their results using such concepts as ligand field strength,
Racah parameters for describing electron repulsion, and the nephelauxetic
effect, to mention just a few terms. Nevertheless LFT has several
severe limitations. For example, LFT can provide only very limited
quantitative indications regarding the molecular structure of a complex
because the application of LFT typically involves first making some
symmetry ---~and hence structural~--- assumptions about the complex.
It is thus very difficult to obtain a detailed description of subtle
variations in the spatial arrangement of ligands around the metal
or to include environmental effects. Despite these difficulties, LFT
is unavoidable, embodying as it does several decades of accumulated
experimental data.

The third Pauling point is given by \emph{ab initio} calculations.
The Latin term \emph{ab initio} means ``from the beginning'' and
means in the present context ``from first principles.'' It most
often designates calculations based upon a single Slater determinant
(Hartree-Fock) or on several Slater determinants carried out in a
self-consistent manner, as for example the complete active space self-consistent
field (CASSCF) method \citep{RTS80}. The self-consistent field (SCF)
calculations must be supplemented with post-SCF calculations in order
to recover important residual dynamic electron correlation effects.
High-level coupled-cluster methods are dynamic correlation methods
able to give results in the complete basis set limit accurate to within
1~kcal~mol$^{-1}$ (350~cm$^{-1}$), the so-called chemical accuracy
\citep{RevModPhys_79_291,NICSeries_42_77}. Their disadvantage is
that the computational resources required (their ``computational
cost'') grows extremely very rapidly with molecular size. For this
reason, they are mainly applied to small transition metal complexes
to obtain important reference data for assessing the performance of
simpler methods such as DFT or of other \emph{ab initio} methods \citep{JChemTheoryComput_8_4216,JChemTheoryComput_10_2306}.
Other \emph{ab initio} methods used in studying spin-crossover complexes
are CASPT2 (which is CASSCF plus second-order perturbation theory
\citep{AMR+90,AMR92}) and spectroscopy-oriented configuration interaction
(SORCI) methods \citep{N03,MCCM93,MDC92}.  The computational cost
of the CASSCF/CASPT2 method limited the reliability of pioneering
work on spin-crossover complexes \citep{B98}. Thanks to  advances
in CASSCF methodology, much more reliable  CASPT2 calculations have
been reported; see Refs~\citep{PV06,PV08,KRLD09,KRL09}, for instance.
Nevertheless we continue to remain pessimistic regarding rigorous
\emph{ab initio} treatment of molecular systems larger than, say,
about 100 atoms. Also, for small molecules, a chemical accuracy better
than~5 kcal~mol$^{-1}$ (1750~cm$^{-1}$) is quite difficult 
to obtain once the molecule contains a transition metal atom. We call
this level of precision the ``best expected accuracy for transition
metal systems'' (BEA-TMS). In what follows, it may be useful to keep
in mind that BEA-TMS = 1750~cm$^{-1}$! Meeting this level of accuracy
is so challenging that certain \emph{ab initio} methods go so far
as to include some semi-empiricism in order to approach it \citep{KRL09,JChemTheoryComput_14_2446,JChemTheoryComput_15_1492}.

The second Pauling point is the method which probably comes closest
at the present time to Artistotle's Golden Mean. This is density-functional
theory (DFT; John Pople and Walter Kohn, 1998 chemistry Nobel prize).
DFT has had several important predecessors \citep{T27,F27,D30,S74}.
Its modern formalism is based upon the two Hohenberg-Kohn theorems
\citep{HK64}, the first of which states that the Hamiltonian of the
molecule is determined up to an arbitrary additive constant by the
ground-state charge density $\rho$. The second Hohenberg-Kohn theorem
establishes that the ground-state energy and charge density may be
determined by minimizing a functional (that is, a function of a function)
of the density.

Although there is an explicit exact constructive formula for this
functional \citep{L79}, it is also impractical and so approximations
must be used in practice. This is why by far the most practical applications
of DFT are carried out using the Kohn-Sham formalism \citep{KS65}
where the density is expressed in terms of a set of self-consistent
Kohn-Sham molecular orbitals. In Kohn-Sham DFT, the key quantity is
the exchange-correlation (xc) energy, $E_{\mathrm{xc}}[\rho]$, which
is a functional of the charge density $\rho$ and is intended to describe
missing exchange, correlation, and kinetic energy contributions which
would otherwise be missing in an exact treatment. Again, as no practical
exact form of the $E_{\mathrm{xc}}$ functional is known, it must
be approximated in practice.

The starting point of nearly every approximation for $E_{\mathrm{xc}}$
is the homogeneous electron gas (HEG or ``jellium'') for which very
precise analytic forms are known for the xc-energy per particle. The
local density approximation (LDA) consists of the assumption that
the xc-energy of a system with an inhomogeneous density is locally
the same as that in a HEG of the same charge density as at that point
of the inhomogeneous system. This approximation works surprisingly
well for optimizing molecular geometries and calculating vibrational
frequencies. But the LDA tends to overestimate bond energies and to
underestimate bond lengths. One way to go beyond the LDA is to include
a dependence on the gradient of the density in the form of the xc-energy
functional through what are known as generalized gradient approximations
(GGAs). These improve both bond energies and bond lengths. Unfortunately
it is difficult to do better within the framework of the pure Kohn-Sham
formalism where the only functional dependence permitted for the xc-energy
functional is on the density. The present tendency is to extend the
functional dependence of the xc-energy functional also to Kohn-Sham
orbitals which are themselves implicit functionals of the charge density.
Perdew has summarized the present hierarchy of functionals in a Jacob's
ladder \citep{PS01,PRC+09} shown in Table~\ref{tab:jacob_s_ladder}.
Climbing \emph{up} the ladder allows the designer of new functionals
more flexibility in designing the xc-energy functional and hence more
degrees of freedom to use to obtain better agreement between theory
and experiment. But this better agreement is by no means guaranteed
(though this is the observed tendency). All that is really guaranteed
is that the calculations will become systematically more expensive.
It is thus important to be able to also climb \emph{down} Jacob's
ladder. For more information about DFT, the authors recommend the
following books in order of increasing difficulty \citep{KH00,PY89,DG90}.

\begin{table}[!htb]
\caption{\label{tab:jacob_s_ladder}Jacob's ladder of density functionals \citep{PS01,C09}
(an updated version is given in Ref.~\citep{PRC+09}; adapted from
Ref.\citep{C09})}

\begin{centering}
{\small{}}%
\begin{tabular}{c|c|c}
\multicolumn{3}{c}{Quantum Chemistry Heaven}\tabularnewline
\hline 
\hline 
 &  & \tabularnewline
{\small{}double-hybrid} & {\small{}-----} & {\small{}$\rho({\bf r}),x({\bf r}),\tau({\bf r}),\psi_{i}({\bf r}),\psi_{a}({\bf r})^{h}$}\tabularnewline
{\small{}hybrid} & {\small{}-----} & {\small{}$\rho({\bf r}),x({\bf r}),\tau({\bf r}),\psi_{i}({\bf r})^{g}$}\tabularnewline
{\small{}mGGA$^{c}$} & {\small{}-----} & {\small{}$\rho({\bf r}),x({\bf r}),\tau({\bf r})^{e}$, $\nabla^{2}\rho({\bf r})^{f}$}\tabularnewline
{\small{}GGA$^{b}$} & {\small{}-----} & {\small{}$\rho({\bf r}),x({\bf r})^{d}$}\tabularnewline
{\small{}LDA$^{a}$} & {\small{}-----} & {\small{}$\rho({\bf r})$}\tabularnewline
 &  & \tabularnewline
\hline 
\hline 
\multicolumn{3}{c}{{\small{}Hartree Model World}}\tabularnewline
\hline 
\end{tabular}{\small\par}
\par\end{centering}
{\small{}$^{a}$ LDA, see text. $^{b}$ GGA, see text. $^{c}$ Meta
GGA. $^{d}$ The reduced gradient $x({\bf r})=\vert\vec{\nabla}\rho({\bf r})\vert/\rho^{4/3}({\bf r})$.
$^{e}$ The local kinetic energy density $\tau({\bf r})=\sum_{p}n_{p}\psi_{p}({\bf r})\nabla^{2}\psi_{p}({\bf r})$.
$^{f}$ There are indications that the local kinetic energy density,
$\tau({\bf r})$, and the Laplacian of the charge density, $\nabla^{2}\rho({\bf r})$,
contain similar information~\citep{PC07}. $^{g}$Occupied orbitals.
$^{h}$Unoccupied orbitals.}{\small\par}
\end{table}

As originally formulated DFT is limited to the study of the ground-state
time-independent properties. The best one might hope within the conventional
formulation is that it will apply to the lowest state of each space
and spin symmetry representation, although present-day xc-functionals
do not have the symmetry dependence that would be rigorously required
by such a generalization \citep{G00,G05,K07}. Time-dependent DFT
(TD-DFT) is intended to treat time-dependent problems and also excited
states. It is based upon theorems analogous to those of Hohenberg
and Kohn and has an associated time-dependent Kohn-Sham equation \citep{RG84}.
Once again the xc-functional is unknown and so must be approximated.
One approximation which is made so often as to virtually define conventional
TD-DFT is to assume that the self-consistant field in the TD Kohn-Sham
equation acts instantaneously and without memory to any time-dependent
change in the charge density. This so-called adiabatic approximation
allows us to use functionals developed for ordinary ground-state static
DFT as xc-functionals in TD-DFT. Excited states may be accessed by
examining the response of the ground state to a time-dependent perturbation.
Response theory then leads to the equation, 
\begin{eqnarray}
\left[\begin{array}{cc}
{\bf A}(\omega_{I}) & {\bf B}(\omega_{I})\\
{\bf B}^{*}(\omega_{I}) & {\bf A}^{*}(\omega_{I})
\end{array}\right]\left(\begin{array}{c}
\vec{X}_{I}\\
\vec{Y}_{I}
\end{array}\right)=\omega_{I}\left[\begin{array}{cc}
{\bf 1} & {\bf 0}\\
{\bf 0} & -{\bf 1}
\end{array}\right]\left(\begin{array}{c}
\vec{X}_{I}\\
\vec{Y}_{I}
\end{array}\right)\,,\label{eq:method.5}
\end{eqnarray}
whose solution gives molecular electronic absorption spectra \citep{C95}.
Here the $\omega_{I}$ (or more precisely $\hbar\omega_{I}$) are
the vertical excitation energies and the corresponding oscillator
strengths, $f_{I}$, are calculated from the vectors $\vec{X}_{I}$
and $\vec{Y}_{I}$. For more information regarding TD-DFT, the reader
is refered to pertinant reviews \citep{C09,C95,L01,D03,MG04,DH05}
and books \citep{MS90,MUN+06}.

Finally, let us note that one of the challenges of modelling spin-crossover
systems is to be able to treat larger and larger systems. The ideal
method should be both simple and accurate, but with a computational
difficulty which increases relatively slowly as the system grows.
It is thus encouraging that DFT is particularly well adapted as a
so-called ${\cal O}(N)$ (order-N) method whose computational difficulty
scales in direct proportion to the number of atoms \citep{HA08} and
there is every reason to believe that TD-DFT can be also made ${\cal O}(N)$.

\section{Application of DFT to the Study of the Structure and Energetics of
Transition-Metal Complexes in their High- and Low-Spin States\label{sec:applications:dft}}

This section discusses the performance of DFT for describing the structural
and energetic properties of low-spin (LS) and high-spin (HS) transition
metal complexes. The discussion will be limited to hexacoordinate
Fe(II) ($d^{6}$) complexes. The ground state of such an octahedral
($O_{h}$) complex has, according to ligand field theory, a LS singlet
$^{1}A_{\mathrm{1g}}(t_{2g}^{6})$ state characterized by maximum
spin pairing of the $d$ electrons in the presence of strong-field
ligands, and a HS quintuplet $^{5}T_{2g}(t_{2g}^{4}e_{g}^{2})$ characterized
by minimal $d$-electron pairing in the presence of weak field ligands.
Orbital degeneracy is lifted by structural distortions due to the
Jahn-Teller effect, but these are typically weak for $O_{h}$ HS $d^{6}$
complexes. The performance of DFT for the determination of the geometries
of LS and HS complexes will be examined first, then its performance
for the HS-LS energy difference. I will be seen that the quality of
xc-functionals is particularly important when it comes to determining
the relative energetics of the spin states of these complexes.

\subsection{Performance of DFT for calculating the geometries of LS and HS complexes}

\begin{table}[!tbh]
\caption{\label{tab:geom}Comparison of theoretical and experimental average
Fe-O distances (\AA{}) for the complex {[}Fe(H$_{2}$O)$_{6}${]}$^{2+}$.
The computational method is indicated in the form \textquotedblleft computational
method/basis set used\textquotedblright{} (where the name of the basis
set is the same as that used in the cited article). Exchange-, correlation-,
and exchange-correlation-type functionals are designated, respectively
by x, c, and xc. Functionals are typically named by the abbreviation
for the exchange part followed by the name for the correlation part
(e.g., PB86 is composed of the x-GGA B and the c-GGA P86). Further
information about abbreviations used in describing functionals is
given in Table~\ref{tab:jacob_s_ladder} and the associated discussion
in Sec.~\ref{sec:methodology}}

\begin{centering}
{\small{}}%
\begin{tabular}{lccc}
\hline 
{\small{}Method} & {\small{}$r_{\mathrm{LS}}$} & {\small{}$r_{\mathrm{HS}}$} & {\small{}$\Delta r_{\mathrm{HL}}$}\tabularnewline
\hline 
{\small{}LDA/A'$^{a}$:1965$^{A}$ (x-LDA) \citep{KS65}} & {\small{}1.912} & {\small{}2.070} & {\small{}0.158}\tabularnewline
{\small{}BP86/E$^{a}$: 1988 (x-GGA) \citep{B88} + 1986 (c-GGA) \citep{P86}} & {\small{}1.996} & {\small{}2.145} & {\small{}0.149}\tabularnewline
{\small{}BLYP/E$^{a}$:1988 (x-GGA) \citep{B88} + 1988 (c-GGA) \citep{LYP88}} & {\small{}2.025} & {\small{}2.166} & {\small{}0.141}\tabularnewline
{\small{}PW91/E$^{a}$: 1991 (xc-GGA) \citep{PBW96}$^{B}$} & {\small{}1.992} & {\small{}2.137} & {\small{}0.145}\tabularnewline
{\small{}PBE/C"$^{a}$: 1996 (xc-GGA) \citep{PBE96}} & {\small{}2.011} & {\small{}2.154} & {\small{}0.143}\tabularnewline
{\small{}RPBE/C"$^{a}$: 1999 (x-GGA) \citep{HHN99} + 1996 (c-GGA)
\citep{PBE96}} & {\small{}2.048} & {\small{}2.187} & {\small{}0.139}\tabularnewline
{\small{}B3LYP/E$^{a}$: 1994 (xc-hybrid)$^{C}$} & {\small{}2.018} & {\small{}2.154} & {\small{}0.136}\tabularnewline
{\small{}Hartree-Fock$^{b}$} & {\small{}2.080} & {\small{}2.193} & {\small{}0.113}\tabularnewline
{\small{}CASSCF(10,12)/II$^{b}$} & {\small{}1.989} & {\small{}2.113} & {\small{}0.124}\tabularnewline
{\small{}CASSCF(12,10)$^{a}$} & {\small{}2.077} & {\small{}2.186} & {\small{}0.119}\tabularnewline
{\small{}Exp.$^{a}$} & {\small{}---} & {\small{}2.068-2.156} & {\small{}---}\tabularnewline
{\small{}LFT$^{c}$} & {\small{}---} & {\small{}---} & {\small{}0.126}\tabularnewline
\hline 
\end{tabular}{\small\par}
\par\end{centering}
{\small{}$^{A}$ VWN parameterization of the c-LDA functional \citep{VWN80}.
$^{B}$ The PW91 functional was first presented in a DFT conference
and then became rapidly incorporated into many DFT codes through sharing
of an implementation of the functional as a }\textsc{\small{}FORTRAN}{\small{}
subroutine. Interestingly no official publication describing the functional
appeared for quite a while. $^{C}$ The B3LYP functional is the same
as the hybrid functioal proposed by Becke \citep{B93}, except that
the c-GGA P91 functional has been arbitrarily replaced by the c-GGA
LYP \citep{LYP88} functional without any reoptimization of parameters.
$^{a}$ Results taken from Ref.~\citep{FMC+04}. $^{b}$ Ref.~\citep{PV06}.
$^{c}$ This LFT estimate is described in Ref.~\citep{FCL+05}.}{\small\par}
\end{table}

Geometries obtained with DFT for LS and HS transition metal complexes
are generally rather good. In order to discuss this concretely, let
us take the case of the complex {[}Fe(H$_{2}$O)$_{6}${]}$^{2+}$.
This complex is referred to as HS because the HS electronic ground
state, which is entropically favored, remains the only populated state
at all temperatures. Of course, the metastable LS state can be examined
theoretically. Geometry optimizations of the HS state with different
functionals lead to a nondegenerate quintuplet with a geometry which
is only slightly distorted due the Jahn-Teller effect. For the LS
state, these optimizations lead to a nondegenerate state of $O_{h}$-symmetry.
In order to describe these structures, the focus will be put on the
mean Fe-O bond length which is given in Table~\ref{tab:geom} for
the HS ($r_{\mathrm{HS}}$) and LS ($r_{\mathrm{LS}}$) states. As
{[}Fe(H$_{2}$O)$_{6}${]}$^{2+}$ is a HS species, only the structure
of the HS state has been characterized experimentally, both in solution
and in crystals. The experimental values given in Table~\ref{tab:geom}
vary by almost 0.09~\AA{}: this great variation emphasizes the importance
of environmental effects on the properties (structural in this case)
of the complexes. There is a large lengthening of the metal-ligand
bond distance after the LS $\rightarrow$ HS change of state as the
result of the promotion of two electrons from the nonbonding $t_{2g}$
orbitals to the antibonding $e_{g}$ orbitals.

The absence of experimental data for LS {[}Fe(H$_{2}$O)$_{6}${]}$^{2+}$
and the fact that the calculations are for gas phase molecules while
experimental measurements are for molecules in condensed phases means
that it is difficult to know which of the predicted $r_{\mathrm{LS}}$,
$r_{\mathrm{HS}}$, and $r_{\mathrm{HL}}$ values should be taken
as most reliable. Nevertheless all values are in good agreement with
what is known experimentally, especially if we take into account the
probable compression of the molecule in the condensed phase. This
level of agreement confirms the earlier claim that one of the strengths
of DFT is the high quality of optimized geometries, which allows one
to go well beyond simple LFT. The performance of different types of
functionals can also be compared. The LDA tends to overestimate the
bond energy. This is seen here by the fact that, compared to other
functionals, the LDA is giving Fe-O bond distances which are too short.
Consistent with their historical purpose, the GGAs correct the overbinding
of the LDA and give longer bonds. Hybrid functionals incorporate a
contribution of HF exchange into the GGA form. As represented here
by the hybrid functional B3LYP, hybrid functionals yield geometries
which are similar to those obtained with GGAs.

Although different functionals give similar HS and LS optimized geometries,
there are important variations of up to 40\% in the difference $\Delta$$r_{\mathrm{HL}}$.
Generally speaking it is difficult to say which functional is best
for calculating geometries, either because the theoretical calculations
are of gas-phase geometries which cannot be directly compared with
condensed-phase measurements or because the molecules are too large
to optimize using high-quality \emph{ab initio} methods. In the specific
case of {[}Fe(H$_{2}$O)$_{6}${]}$^{2+}$, if ones believes the LFT
estimation for $\Delta$$r_{\mathrm{HL}}$(which is based upon decades
of empirical experience) and the CASSCF calculations, then the HF
results (which lack electron correlation) and LDA calculations can
be ruled out in favor of DFT calculations with GGAs or hybrid functionals.
This conclusion is also expected to hold for most transition metal
complexes.

\subsection{Performance of DFT for spin-crossover energies}

When considering the HS-LS energy difference, one needs to distinguish
between the HS-LS zero-point energy difference, $\Delta E_{\mathrm{HL}}^{\circ}=E_{\mathrm{HS}}^{\circ}-E_{\mathrm{LS}}^{\circ}$,
its electronic contribution, $\Delta E_{\mathrm{HL}}=E_{\mathrm{HS}}-E_{\mathrm{LS}}$,
and its vibrational contribution, $\Delta E_{\mathrm{HL}}^{\mathrm{v}}=E_{\mathrm{HS}}^{\mathrm{v}}-E_{\mathrm{LS}}^{\mathrm{v}}$
($\Delta E_{\mathrm{HL}}^{\circ}=\Delta E_{\mathrm{HL}}+\Delta E_{\mathrm{HL}}^{\mathrm{v}}$).
$\Delta E_{\mathrm{HL}}$ can be obtained by optimizing HS and LS
geometries, while $\Delta E_{\mathrm{HL}}^{\circ}$ requires additional
vibrational frequency calculations for the HS and LS complexes. Density-functional
calculations with GGA or hybrid functionals typically give very good
frequencies for complexes and hence good estimates of $\Delta E_{\mathrm{HL}}^{\mathrm{v}}$.
The situation is more complicated when it comes to determining $\Delta E_{\mathrm{HL}}$
which, in spin-crossover complexes, takes values on the order of 100
$\cong$ 1000 cm$^{-1}$. For such calculations, the best \emph{ab
initio} methods have a BEA-TMS of around 1750 cm$^{-1}$ (Sec.~\ref{sec:methodology}).
This is thus quite a challenge for DFT.

\begin{table}
\caption{\label{tab:DeltaEHB_2_EN}Comparison of the energy difference $\Delta E_{\mathrm{HL}}=E_{\mathrm{HS}}-E_{\mathrm{LS}}$
for different functionals (cm$^{-1}$) for well-characterized iron(II)
complexes. The spin of the ground state is given in parentheses. For
the $D_{3}$ complex {[}Fe(bpy)$_{3}${]}$^{2+}$ (bpy = 2,2'-bipyridine),
note that the $^{5}E$ and $^{5}A_{1}$ are quasidegenerate~\citep{LVH+05}
and correspond to the $^{5}T_{2g}$ LFT state. We take the HS LFT
state to be the lower energy $^{5}A_{1}$ state in this case. The
functionals are listed in chronological order}

\begin{centering}
{\small{}}%
\begin{tabular}{lccc}
\hline 
\multicolumn{1}{c}{{\small{}Method}} & \multicolumn{3}{c}{{\small{}Molecule}}\tabularnewline
{\small{}functional: year (type) {[}reference{]}} & {\small{}{[}Fe(H$_{2}$O)$_{6}${]}$^{2+}$ (HS)} & {\small{}{[}Fe(NH$_{3}$)$_{6}${]}$^{2+}$ (HS)} & {\small{}{[}Fe(bpy)$_{3}${]}$^{2+}$ (LS)}\tabularnewline
\hline 
{\small{}LDA: 1965$^{A}$ (Local-x) \citep{KS65}} & {\small{}$-$3 896$^{a}$} & {\small{}8 187$^{b}$} & \tabularnewline
{\small{}BP86: 1988 (GGA-x) \citep{B88}} &  &  & \tabularnewline
{\small{}$\,\,\,\,\,\,\,\,$ + 1986 (GGA-c) \citep{P86}} & {\small{}$-$8798$^{a}$} & {\small{}790$^{b}$} & {\small{}11135$^{d}$}\tabularnewline
{\small{}PW91: 1991 (GGA-xc) \citep{PBW96}$^{B}$} & {\small{}-9232$^{a}$} & {\small{}617$^{b}$} & {\small{}11887$^{d}$}\tabularnewline
{\small{}B3LYP: 1994 (Hybrid-xc){[}$,${]}$^{C}$} & {\small{}$-$11465$^{a}$} & {\small{}$-$4 978$^{b}$} & {\small{}1211$^{d}$}\tabularnewline
{\small{}PBE: 1996 (GGA-xc) \citep{PBE96}} & {\small{}$-$10081$^{a}$} & {\small{}$-$147$^{b}$} & {\small{}11337$^{d}$}\tabularnewline
{\small{}HCTH93: 1998 (GGA-xc) \citep{HCTH98}} & {\small{}$-$18779$^{b}$} & {\small{}$-$9430$^{b}$} & \tabularnewline
{\small{}HCTH147: 1998 (GGA-xc) \citep{HCTH98}} & {\small{}$-$18211$^{b}$} & {\small{}$-$8576$^{b}$} & \tabularnewline
{\small{}HCTH407: 1998 (GGA-xc) \citep{HCTH98}} & {\small{}$-$19631$^{b}$} & {\small{}$-$10682$^{b}$} & \tabularnewline
{\small{}VSXC: 1998 (meta-GGA-xc) \citep{VS98}} & {\small{}$-$13975$^{e,f}$} & {\small{}$-$5928$^{e,f}$} & {\small{}$-$1822$^{e,f}$}\tabularnewline
{\small{}RPBE: 1999 (GGA-xc) \citep{HHN99}} & {\small{}$-$11844$^{a}$} & {\small{}$-$2744$^{b}$} & {\small{}6640$^{d}$}\tabularnewline
{\small{}PBE0$^{D}$: 1999 (Hybrid-xc) \citep{AB99}} & {\small{}$-$14504$^{b}$} & {\small{}$-$7195$^{b}$} & {\small{}$-$466$^{d}$}\tabularnewline
{\small{}OLYP$^{E}$: 2001 (GGA-x) \citep{HC01}} &  &  & \tabularnewline
{\small{}$\,\,\,\,\,\,\,\,$ + 1988 (GGA-c) \citep{LYP88}} & {\small{}$-$16284$^{e,f}$} & {\small{}$-$6693$^{e,f}$} & {\small{}2684$^{e,f}$}\tabularnewline
{\small{}OPBE$^{E}$: 2001 (GGA-x) \citep{HC01}} &  &  & \tabularnewline
{\small{}$\,\,\,\,\,\,\,\,$ + 1988 (GGA-c) \citep{PBE96}} & {\small{}$-$17200$^{g}$} & {\small{}$-$6650$^{g}$} & {\small{}5210$^{g}$}\tabularnewline
{\small{}B3LYP$^{\star}$: 2002 (Hybrid-xc) \citep{SRH02}} & {\small{}$-$10456$^{e,f}$} & {\small{}$-$3226$^{e,f}$} & {\small{}3332$^{e,f}$}\tabularnewline
\multicolumn{4}{l}{Best Estimates}\tabularnewline
{\small{}Exp.} &  &  & {\small{}3500 --- 6000$^{d}$}\tabularnewline
\emph{\small{}Ab initio}{\small{}$^{c}$} & {\small{}$-$13360 --- $-$12347$^{a}$} & {\small{}$-$11230 --- $-$9125$^{b}$} & \tabularnewline
{\small{}Average} & {\small{}$-$12853} & {\small{}$-$10177} & {\small{}4750}\tabularnewline
\hline 
\end{tabular}{\small\par}
\par\end{centering}
{\small{}$^{A\mbox{--}C}$ See Table~\ref{tab:geom}. $^{D}$ Also
known as PBE1PBE. $^{E}$ The OPTX GGA-x \citep{HC01} is often designated
by the letter O, as in the xc-functionals OLYP and OPBE. $^{a}$ Ref.~\citep{FMC+04}.
$^{b}$ Ref.~\citep{FCL+05}. $^{c}$ Results of CASPT2 and SORCI
calculations with empirical atomic energy correction \citep{FMC+04,FCL+05}.
$^{d}$ Ref.~\citep{LVH+05}. $^{e}$ Ref.~\citep{GBF+05}. $^{f}$
At the geometry optimized using the same functional for these complexes.
Note that the choice of functional has relatively little effect on
the geometry optimizations. $^{g}$ Ref.~\citep{S07}.}{\small\par}
\end{table}

The current situation may be summed up in the following way. Hartree-Fock
tends to strongly underestimate $\Delta E_{\mathrm{HL}}$. The LDA
functional tends to strongly overestimate $\Delta E_{\mathrm{HL}}$.
The GGAs tend to correct the problems with the LDA. But DFT calculations
prior to the beginning of this century tended to overestimate the
stability of the LS state with respect to the HS state. The same studies
showed that the B3LYP hybrid functional, which was the most sophisticated
widely-used functional at that time, overestimated the stability of
the HS state with respect to the LS state. Thus $\Delta E_{\mathrm{HL}}$
is underestimated as with the HF method. The beginning of the 2000s
saw a significant increase in the performance of density functionals.
One example was the B3LYP$^{\star}$ functional obtained by empirically
decreasing the amount of HF exchange in the B3LYP functional from
20\% to 15\% so as to reproduce the ground states of a series of complexes
\citep{SRH02}. It is important to be able to descend Jacob's ladder
(Sec.~\ref{sec:methodology}), and the advent of new and more sophisticated
GGAs seemed to offer interesting new possibilities. In fact, when
it came to testing the ability of these functionals for calculating
$\Delta E_{\mathrm{HL}}$, the biggest problem was to collect and
verify reliable comparison data. This is far from trivial because
electronic structure calculations are usually made for the isolated
complex while experimental data is often for complexes in condensed
media and hence have a different enthalpy and free energy dependence
than in the gas phase. For example, Table~\ref{tab:DeltaEHB_2_EN}
gives results obtained for the values of $\Delta E_{\mathrm{HL}}$
for the complexes {[}Fe(H$_{2}$O)$_{6}${]}$^{2+}$, {[}Fe(NH$_{3}$)$_{6}${]}$^{2+}$,
and {[}Fe(bpy)$_{3}${]}$^{2+}$ (bpy = 2,2'-bipyridine). Quality
benchmarks have been obtained for the small HS complexes {[}Fe(H$_{2}$O)$_{6}${]}$^{2+}$
\citep{FMC+04} and {[}Fe(NH$_{3}$)$_{6}${]}$^{2+}$ by high-level
\emph{ab initio} CASPT2 and SORCI calculations \citep{FMC+04,FCL+05}.
As to the LS complex {[}Fe(bpy)$_{3}${]}$^{2+}$, a reliable estimation
of $\Delta E_{\mathrm{HL}}$ has been obtained by an experimental
study of the HS $\rightarrow$ LS relaxation rate of the complex in
different solid matrices \citep{LVH+05,HEL+06}.

The uncertainty of the \emph{ab initio} results is 1013 cm$^{-1}$
for the {[}Fe(H$_{2}$O)$_{6}${]}$^{2+}$ complex and 2105 cm$^{-1}$
for the {[}Fe(NH$_{3}$)$_{6}${]}$^{2+}$ complex, which is consistent
with the BEA-TMS of 1750 cm$^{-1}$. The uncertainty of the experimental
value found for the {[}Fe(bpy)$_{3}${]}$^{2+}$ complex is larger
(2~500 cm$^{-1}$), reflecting the influence of the environment on
$\Delta E_{\mathrm{HL}}$. The LDA and the GGAs before 1998 (BP86,
PW91, RPBE) overestimate the difference $\Delta E_{\mathrm{HL}}$.
Also, for {[}Fe(bpy)$_{3}${]}$^{2+}$, $\Delta E_{\mathrm{HL}}$
is underestimated for the classic hybrids B3LYP and PBE0. This is
different than the case of the {[}Fe(H$_{2}$O)$_{6}${]}$^{2+}$
complex for which $\Delta E_{\mathrm{HL}}$ is underestimated by PBE0
but which is nevertheless given reasonably well by the B3LYP functional.
Finally, the two hybrid functionals both overestimate the best estimate
of $\Delta E_{\mathrm{HL}}$ for {[}Fe(NH$_{3}$)$_{6}${]}$^{2+}$.
The reparameterized functional B3LYP$^{\star}$ gives a distinct improvement
over the original B3LYP functional for $\Delta E_{\mathrm{HL}}$ for
{[}Fe(bpy)$_{3}${]}$^{2+}$. On the other hand, B3LYP$^{\star}$
is even worse than B3LYP in so far as B3LYP$^{\star}$ results in
a greater overestimation of $\Delta E_{\mathrm{HL}}$ than does B3LYP
for the {[}Fe(H$_{2}$O)$_{6}${]}$^{2+}$ and {[}Fe(NH$_{3}$)$_{6}${]}$^{2+}$
complexes which are, after all, chemically very different than the
complexes originally used to parameterize the B3LYP$^{\star}$ functional.
The post-1996 functionals such as RPBE, the ``HCTH'' functionals,
and the OLYP and OPBE functionals give better results than previous
GGAs. The OLYP and OPBE functionals differ only by the correlation
functionals whose choice is less important for the relative spin energies
than is that of the exchange functional. At the present time, these
two GGAs are two of the best performing functionals for iron(II) complex
spin-state energetics. Thus, the OLYP gives better results than B3LYP$^{\star}$
for the calculation of free energy differences $\Delta G_{\mathrm{HL}}$
in the series originally used to parameterize the B3LYP$^{\star}$
functional \citep{GBF+05}. In fact, it would appear that the contribution
of HF exchange in the B3LYP$^{\star}$ functional should be further
decreased below its current value of 15\% in order to improve the
performance of this hybrid \citep{R02,GBF+05}. The dependence of
$\Delta E_{\mathrm{HL}}$ with respect to the contribution of HF exchange
has also been studied in the case of {[}Fe(bpy)$_{3}${]}$^{2+}$.
The results in this latter case suggest that the proportion of HF
exchange should be reduced to around 10\% not just in the B3LYP functional
but also in the PBE0 functional \citep{LVH+05}. Also note that a
 study has shown the OPBE functional to perform well for spin-state
energetics in iron(II) and iron(III) complexes \citep{S07}.

It would seem that in going from the B3LYP$^{\star}$ to the OLYP
and OPBE GGAs that Jacob's ladder has been successfully descended
without any noticeable loss of precision. One may thus envisage with
confidence studying much larger transition metal compounds than those
studied here. Further improvement of spin-state energetics is likely
to continue as long as work continues to design better and better
density functionals.

A major problem linked to the use of DFT for spin-state energetics
is that it is difficult and risky to anticipate the performance of
a functional for spin-crossover complexes on the basis of its good~--
even excellent~-- performance for other complexes. It would thus
be interesting to be able to estimate $\Delta E_{\mathrm{HL}}$ by
somehow avoiding the functional dependence. Thus, supposing that the
different xc-functionals give very similar potential energy surfaces
which are nevertheless shifted in energy with respect to each other,
it has been shown \citep{ZBF+07} that the difference $\Delta(\Delta E_{\mathrm{HL}})$
of the $\Delta E_{\mathrm{HL}}$ energy differences for the two complexes
C$_{1}$ and C$_{2}$, $\Delta(\Delta E_{\mathrm{HL}})=\Delta E_{\mathrm{HL}}(\mbox{C\ensuremath{_{1}}})-\Delta E_{\mathrm{HL}}(\mbox{C\ensuremath{_{2}}})$,
should be roughly independent of the choice of functional. This hypothesis
has been verified by calculations on two iron(II) spin-transition
complexes using 53 functionals including GGAs, hybrid, and meta-GGAs
\citep{ZBF+07}. The GGAs used gave results in very good agreement
with the hypothesis while the hybrids were in less good agreement
with the hypothesis and the meta-GGAs failed completely. Nevertheless
it seems possible, knowing $\Delta E_{\mathrm{HL}}$ for a reference
complex, to use DFT calculations to estimate its value for other complexes
with a precision similar to, or possibly even better than, the BEA-TMS
\citep{ZBF+07,PhysChemChemPhys_15_3752}.

\section{Excited States from TD-DFT\label{sec:applications:tddft}}

Among the methods for describing molecular excited states, TD-DFT
is probably the one which is currently the most used for studying
large and middle-sized molecules. For reasons of brevity, only a few
observations will be made on the place of TD-DFT in the study of transition
metal complexes, and particularly spin-crossover complexes.

A very important field of application of TD-DFT is modeling the photophysical
properties of ruthenium(II) complexes, which are $d^{6}$ closed-shell
complexes which are of interest for several reasons, including the
fact that they are used as photosensitizers in Grätzel's dye-sensitized
photocells \citep{G01}. In fact, the number of papers reporting TD-DFT
calculations for this particular application has really taken off,
in part, because of the ready availability of the method in the vast
majority of quantum chemistry and quantum physics programs as well
as because of the good agreement obtained between theory and experiment
with this approach.

In fact, as far as the application of TD-DFT to closed shell ruthenium(II)
complexes around their equilibrium geometries, the situation is relatively
simple in the sense that one essentially needs only concern oneself
with the choice of the functional to use.\footnote{An atomic orbital basis set must also be chosen so as to give an adequate
description of the Kohn-Sham molecular orbitals, but the basis set
problem is shared by TD-DFT and excited-state \emph{ab initio} methods.
In general larger basis sets may be used in TD-DFT calculations than
in sophisticated excited-state \emph{ab initio} methods.} Moreover most GGA and hybrid functionals give a good description
of the valence excited states of these molecules. However, the domaine
of validity of TD-DFT for practical calculations is limited by the
approximate nature of these functionals and the use of the TD-DFT
adiabatic approximation (Sec.~\ref{sec:methodology}), TD-DFT is
not without its limitations. For example, the characterization of
high-energy (Rydberg) excited states requires the use of functionals
whose xc-potentials have the correct asymptotic behavior. Also there
are problems describing charge-transfer excitations and those with
significant multiple-excitation character. The reader seeking more
information about the strengths and limitations of TD-DFT and how
the limitations may be attenuated or eliminated altogether are referred
to Refs.~\citep{C95,GK90,GUG94,C96,BG98,L01,VBB+01,C02,D03,MG03,MG04,BWG05,DH05,CMA+06,C08,C09,MS90,MUN+06}
for more detailed information.

\begin{figure}[!tbh]
\begin{centering}
\includegraphics[scale=0.38]{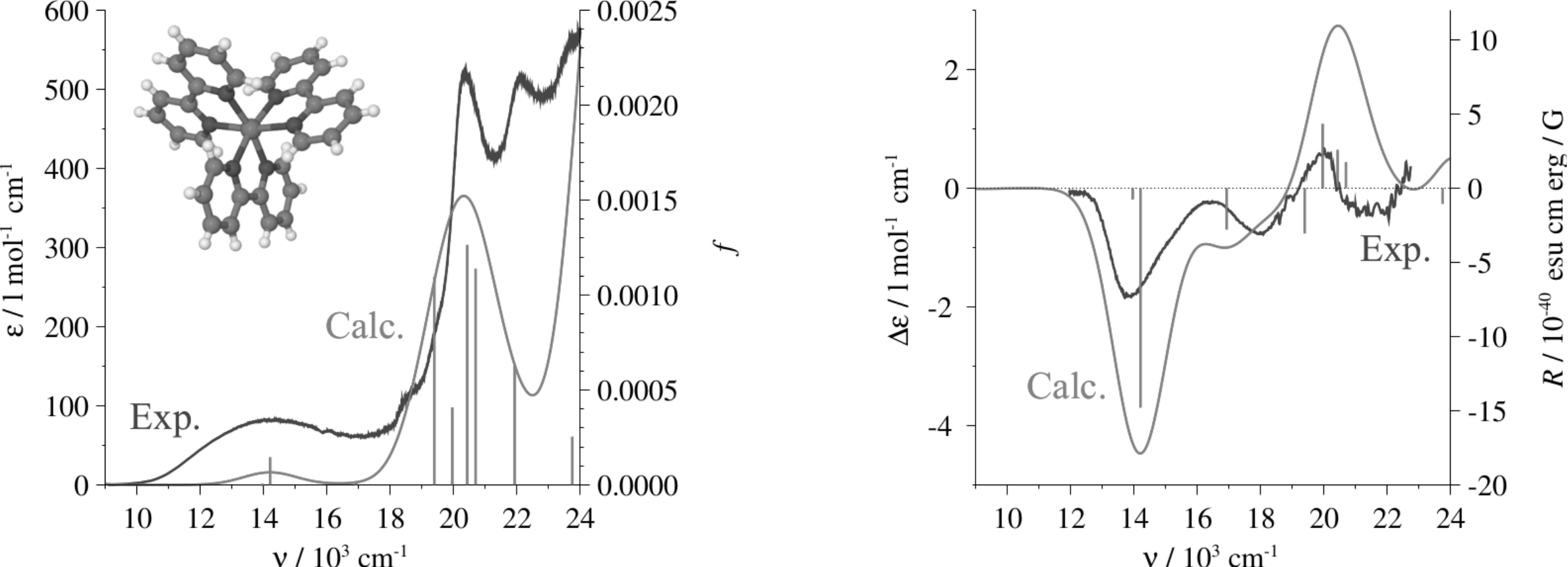}
\par\end{centering}
\caption{Absorption (left) and circular dichroism (right) spectra calculated
for the $\Lambda$ enantiomer of the {[}Co(bpy)$_{3}${]}$^{2+}$
complex in the LS state (TDB3LYP/TZVP calculation.) Also shown are
the measured absorption and dichroism spectra for a single crystal
of {[}Co(bpy)$_{3}${]}{[}LiRh(ox)$_{3}${]} at respectively 11 K
and 15 K. Figure adapted from Ref.~\citep{VZK+06}. \label{fig:CD}}
\end{figure}

TD-DFT may also be used to study the optical properties of open-shell
species such as the $d^{7}$ complex {[}Co(bpy)$_{3}${]}$^{2+}$.
The absorption and circular dichroism spectra of this chiral complex
were calculated at the TD-B3LYP/TZVP level \citep{VZK+06}. Figure~\ref{fig:CD}
shows that the spectra obtained for the $\Lambda$ enantiomer of the
LS complex, while taking into account the Jahn-Teller instability
of this state (see Ref.~\citep{VZK+06}). The calculated oscillator
strengths and rotatory power (see below) are each represented by stick
spectra. To simulate the absorption spectrum $\epsilon=\epsilon(\tilde{\nu})$
($\epsilon$ in M$^{-1}\,$cm$^{-1}$, $\tilde{\nu}$ in cm$^{-1}$),
the molar extinction coefficient $\epsilon_{I}(\tilde{\nu})$ associated
with the $I$th transition centered at $\tilde{\nu_{I}}$ has been
calculated by convoluting the oscillator strength $f_{I}$ with a
gaussian with a full-width at half-height of 2~000 cm$^{-1}$ whose
normalization guarantees the equality, 
\begin{equation}
f_{I}=(\mbox{\ensuremath{4.32\times10^{-9}} M\,cm\ensuremath{^{2}}})\int\epsilon_{I}(\tilde{\nu})\,d\tilde{\nu}\,.\label{eq:tddft.1}
\end{equation}
The rotatory power $R_{I}$ of this transition is related to the difference
$\Delta\epsilon_{I}$ between the absorption coefficients for left-circularly
polarized light ($\epsilon_{g}$) and right-circularly polarized light
($\epsilon_{d}$) ($\Delta\epsilon=\epsilon_{g}-\epsilon_{r}$), by
the relation, 
\begin{equation}
R_{I}=(\mbox{\ensuremath{22.97\times10^{-40}} cgs\,M\,cm})\int_{0}^{\infty}\frac{\Delta\epsilon(\tilde{\nu})}{\tilde{\nu}}\,{\rm d}\tilde{\nu}\,.\label{eq:tddft.2}
\end{equation}
In order to model circular dichroism spectra, $\Delta\epsilon_{I}(\tilde{\nu})$
has been calculated by convoluting the rotatory power $R_{I}$ using
a Gaussian whose full width at half height is 2~000 cm$^{-1}$ and
whose normalization guarantees the above equality. Figure~\ref{fig:CD}
compares the calculated spectra with those measured for a monocrystal
of the spin-crossover compound {[}Co(bpy)$_{3}${]}{[}LiRh(ox)$_{3}${]}
which is LS at the cryogenic temperatures used in the measurement.
The agreement between theory and experiment is quite good. This allows
to identify the configuration of the {[}Co(bpy)$_{3}${]}$^{2+}$
complex as $\Lambda$ \citep{VZK+06}.

Although TD-DFT provides a semiquantitative way to predict the envelope
of the spectrum, it does not necessarily allow to do this by giving
a good description of the underlying states. This depends upon the
quality of the functional used and, for open-shell systems, on the
degree of spin-contamination. The TD-DFT study of open-shell systems
may suffer from complications due to spin contamination in the ground
or in the excited states. Spin contamination occurs when the approximate
wave function $\Psi$ used to describe the $S$ spin state is not
an eigenfunction of the spin operator $\hat{\mathbf{S}}^{2}$ with
corresponding eigenvalue $S(S+1)$. The diagnosis of the degree of
spin contamination is obtained by calculating $\langle\hat{\mathbf{S}}^{2}\rangle=\langle\Psi\left|\mathbf{S}^{2}\right|\Psi\rangle$
and comparing it with the expected value of $S(S+1)$. The problem
of spin contamination is typically less important in DFT than in Hartree-Fock
calculations. In TD-DFT calculations, its complete (or partial) absence
is a necessary, but not sufficient, condition for the correct description
of the excited state.

\begin{figure}[htb]
\centering{}\includegraphics[scale=0.6]{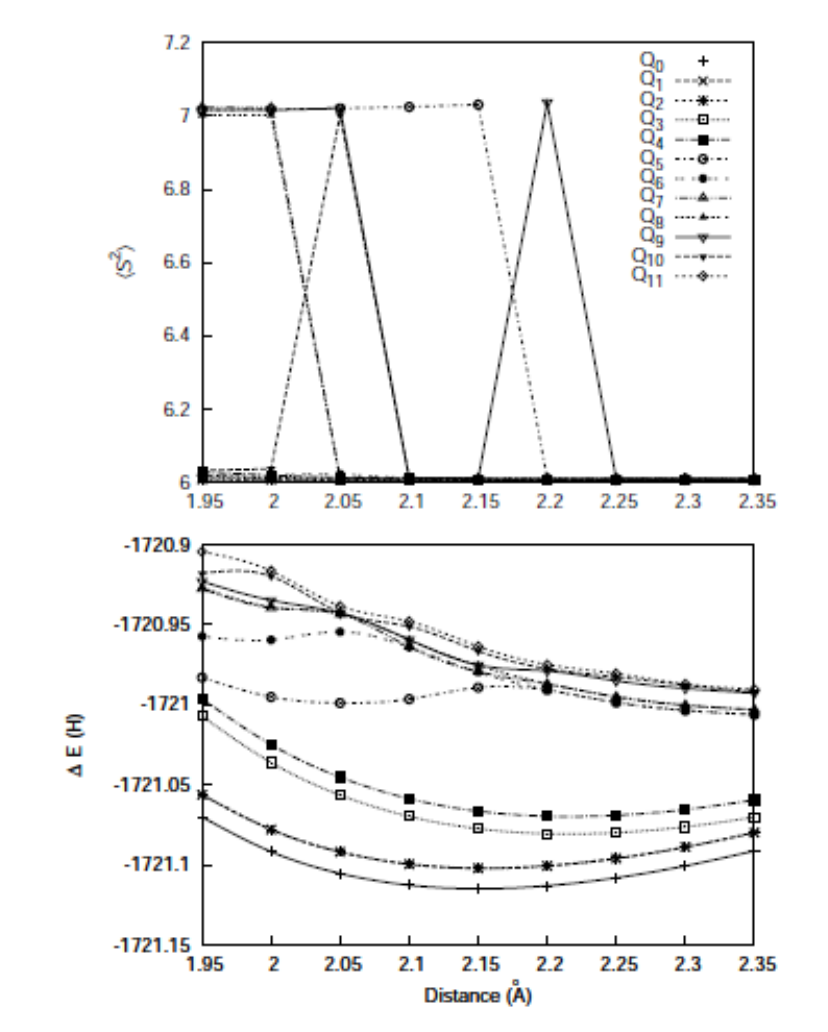} \caption{\label{fig:contam}Potential energy curves for ground ($Q_{0}$) and
excited ($Q_{I\protect\geq1}$) quintuplet states of the {[}Fe(H$_{2}$O)$_{6}${]}$^{2+}$
complex along the Fe-O breathing coordinate {[}$R$(Fe-O){]} and the
corresponding calculated expectation values $\langle\hat{\mathbf{S}}^{2}\rangle$
calculated along the same potential energy curve. The DFT calculations
have been calculated using the PBE functional for the ground state
and using TD-DFT in the TDA and the adiabatic LDA for the excited
states.}
\end{figure}

Thanks to recent work \citep{CIC06,C09,ICJC09,MC17}, there are now
tools which allow to automatically assign the spin of excited sates
in TD-DFT, and thus to identify and eventually correct spin-contamination.
Figure~\ref{fig:contam} illustrates this approach in the case of
the {[}Fe(H$_{2}$O)$_{6}${]}$^{2+}$ complex \citep{F06,F07,B09,J09}.
Potential energy curves are shows for the ground ($Q_{0}$) and excited
($Q_{I\geq1}$) quintuplet states of the {[}Fe(H$_{2}$O)$_{6}${]}$^{2+}$
complex along the Fe-O distance associated with the totally symmetric
breathing mode. The curves for the expectation values $\langle\hat{\mathbf{S}}^{2}\rangle$
calculated at the same geometries are shown. The four lowest energy
curves correspond to the ground and $d\rightarrow d$ type excited
sates. These states are spin-contamination free ($\langle\hat{\mathbf{S}}^{2}\rangle=6$).
The higher energy curves correspond to metal $\rightarrow$ ligand
charge transfer excitations. Depending upon the geometry of the complex
these latter states show spin-contamination ($\langle\hat{\mathbf{S}}^{2}\rangle=7$)
and must either be discarded or corrected to make them physical \citep{ICJC09}.

Spin-contamination is at least partially a consequence of the simplicity
of TD-DFT in its current developmental state. But even this simplicity
is very beneficial as it allows photodynamics modeling using Tully-type
mixed TD-DFT/classical trajectory surface hopping \citep{CJI+07,TTR07,MWB08,TTR+08,WMSB08,BPP+10}.
The application to the photochemistry of oxirane of this technique
for modeling nonadiabatic dynamics has been very instructive \citep{CJI+07,TTR+08}.
It is currently believed that most photoreactions pass either through
or near conical intersections and that nuclear dynamics is often needed
in order to understand photoreaction products. It is hence essential
for TD-DFT to be able to give a good description of systems near conical
intersections. Conventional TD-DFT uses the adiabatic approximation
(i.e., the frequency dependence of the \textbf{A} and \textbf{B} matrices
in Eq.~(\ref{eq:method.5}) is neglected). A seminal TD-DFT study
of oxirane has shown the importance of the Tamm-Dancoff approximation
(TDA) {[}\textbf{B}=0 in Eq.~(\ref{eq:method.5}){]} to reduce the
harmful effects of spin-instabilities in TD-DFT calculations of the
potential energy surfaces \citep{CJI+07}. Later it was seen that
applying mixed TDA TD-DFT/classical trajectory surface hopping gave
results in good agreement with known experimental results and provided
a more complete understanding of the classic Gomer-Noyes mechanism
of the reaction \citep{TTR+08}. This was surprising from the point
of view that conventional TD-DFT cannot provide true conical intersections
between the ground and excited states \citep{LKQM06}. Instead the
potential energy surfaces obtained by TDA TD-DFT and CASSCF have almost
the same appearance near the conical intersection determined by the
CASSCF calculations except that the TDA TD-DFT has two slightly interpenetrating
cones instead of a true conical intersection. This detail may be important
for determining the surface hopping rate. But the results of Ref.~\citep{TTR+08}
show that the TDA TD-DFT description is adequate for predicting the
mechanism of the photoreaction.

\section{Inclusion of Environmental Effects\label{sec:environmental-effects}}

Up to this point the focus has been put on the properties of isolated
complexes. The influence of the environment of the complex is by no
means anodine. Consider the case of hexacoordinated iron(II) complexes.
Under a LS $\rightarrow$ HS transition, their molecular volume increases
considerably due to lengthening of the metal-ligand bonds (Table \ref{tab:geom}).
This large volume change is the origin of the pressure sensitivity
of spin-crossover and of the sensitivity of spin-crossover to modifications
of the second coordination sphere of the complexes \citep{HEL+06}.
Thus the application of an external pressure destabilizes the HS state
with respect to the LS state which is has a smaller molecular volume,
and hence also increases the zero-point energy difference $\Delta E_{\mathrm{HL}}^{\circ}$.
By analogy, the environmental influence on $\Delta E_{\mathrm{HL}}^{\circ}$
can also be rationalized in terms of ``internal'' or ``chemical''
pressure which destabilizes the HS state with respect to the LS state.

The value of $\Delta E_{\mathrm{HL}}^{\circ}$ determines not only
the spin-crossover transition but also the kinetics of the HS $\rightarrow$
LS intersystem crossing which follows the photoinduced population
of the HS state. The HS $\rightarrow$ LS relaxation is observed after
photoexcitation for any LS iron(II) complex, such as typical spin-crossover
systems or {[}Fe(bpy)$_{3}${]}$^{2+}$. At cryogenic temperatures,
relaxation takes place through tunnelling. In so far as, for a chemically
similar family of complexes, the structural change induced by a spin
change may be mainly described as a uniform change in metal-ligand
distances, the relaxation rate constant in the tunnelling regime $k_{{\rm HL}}(T\rightarrow0\,\mathrm{K})$
may be written as a simple function of $\Delta E_{HL}^{\circ}$ and
of the change $\Delta r_{\mathrm{HL}}$ in the average metal-ligand
bond length, $k_{{\rm HL}}(T\rightarrow0\,\mathrm{K})=f(\Delta r_{\mathrm{HL}},\Delta E_{\mathrm{HL}}^{\circ})$
\citep{H04b}. In the case of spin-crossover systems, the parameters
$\Delta r_{\mathrm{HL}}$ and $\Delta E_{\mathrm{HL}}^{\circ}$ may
be determined experimentally: Thus bond-length difference $\Delta r_{\mathrm{HL}}$
may be found by comparing crystal structures for LS and HS complexes.
The energy difference $\Delta E_{HL}^{\circ}$ is proportional to
the spin-crossover temperature $T_{1/2}$.

Thus, for {[}Fe(II)N$_{6}${]} spin-crossover transition complexes,
it is known that $\Delta r_{\mathrm{HL}}$ is about 0.2~Å and $k_{{\rm HL}}(T\rightarrow0\,\mathrm{K})\approx g(\Delta E_{\mathrm{HL}})$
\citep{H04b}. Consequently, it is possible to estimate $\Delta E_{\mathrm{HL}}$
from $k_{{\rm HL}}(T\rightarrow0\,\mathrm{K})$ for any LS {[}Fe(II)N$_{6}${]}-type
spin-crossover complex. This approach has been applied to the LS complex
{[}Fe(bpy)$_{3}${]}$^{2+}$ and it was found that 2~500~cm$^{-1}$
$\leq\Delta E_{\mathrm{HL}}\leq$ 5~000~cm$^{-1}$. To obtain this,
as the crystallographic structure of the HS state of this LS complex
is not known, it was necessary to make the ansatz that $\Delta r_{\mathrm{HL}}\approx0.2$~Å.
However this ansatz was confirmed by DFT calculations \citep{LVH+05}.
Note that this approach is only useful when the spin-crossover-induced
structural change is mainly due to a uniform change of the metal-ligand
bond distances. A notable exception where the structural change is
more complicated is for the {[}Fe(tpy)$_{2}${]}$^{2+}$ complex which,
like the {[}Fe(bpy)$_{2}${]}$^{2+}$ complex, has polypyridine ligands
\citep{HEL+06}.

\begin{figure}[!tbh]
\centering{}\includegraphics[scale=0.6]{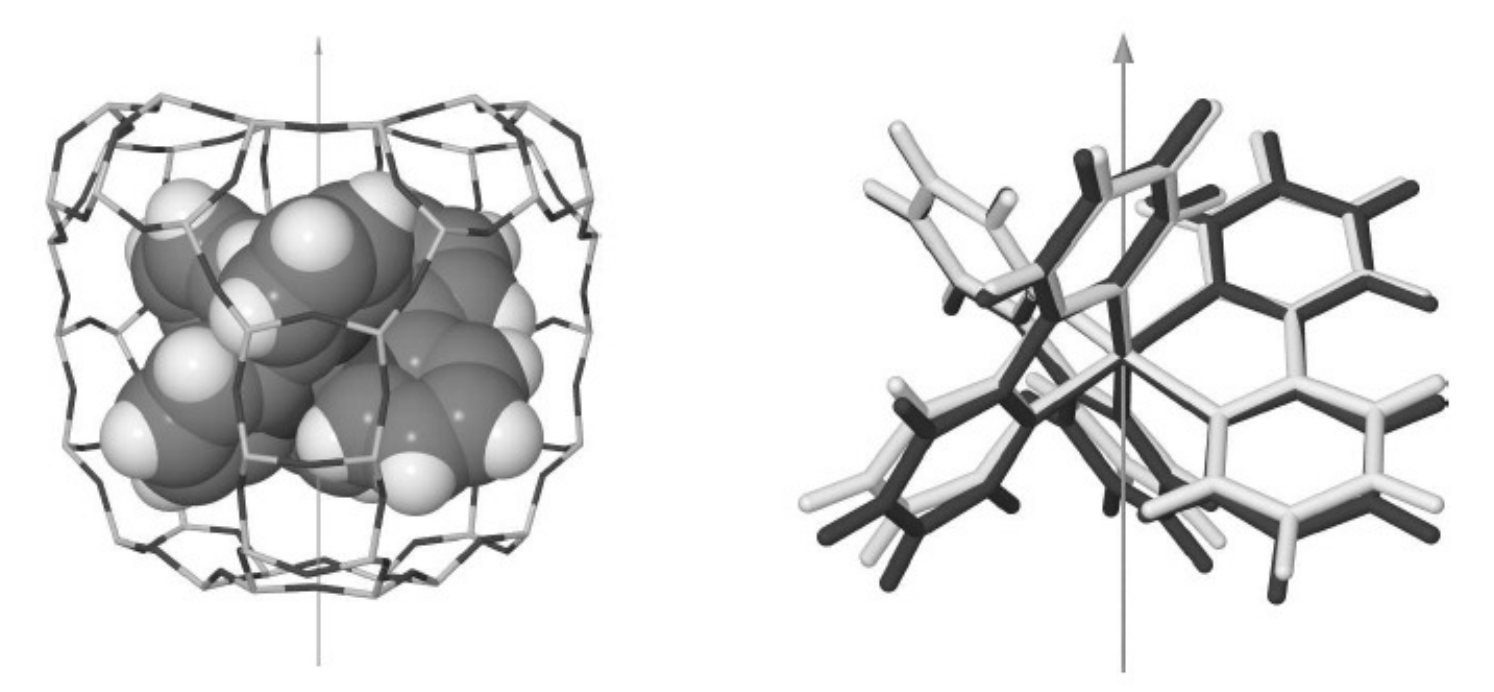} \caption{Left: $C_{3}$ supermolecular model of the inclusion compound {[}Fe(bpy)$_{3}${]}$^{2+}$@Y:
the model is composed of the complex along with the surrounding atoms
of oxygen and silicon which define the supercage. Note that the silicon
atom valence is saturated by hydrogen atoms which are not shown for
graphic simplicity. Right: Comparison of the LS gas phase (black)
and zeolite-embedded (white) {[}Fe(bpy)$_{3}${]}$^{2+}$ geometries.
The cage effect gives a slight contraction and distortion of the geometry
of the complex. Figure reproduced with permission from Ref.~\citep{VHL09}.
\label{fig:LSinY}}
\end{figure}

In fact, though the spin-state energetics are similar in these two
complexes, different values of $k_{{\rm HL}}(T\rightarrow0\,\mathrm{K})$
have been measured in the range $10^{-6}$-$10^{7}$~s$^{-1}$ in
different environments. For {[}Fe(bpy)$_{3}${]}$^{2+}$ doped in
isostructural host crystals {[}M(bpy)$_{3}${]}(PF$_{6}$)$_{2}$
(M = Co, Zn, Mn, Cd), $k_{{\rm HL}}(T\rightarrow0\,\mathrm{K})$ increases
monotonically from $\sim$10$^{3}$ to $\sim$10$^{6}$~s$^{-1}$
with a corresponding shrinking of the unit cell which, over the host
series, is $\sim$2\% \citep{HEL+06}. This example allows to emphasize
the usefulness of the concept of ``chemical pressure'' for explaining
the influence of the environment on the electronic properties of spin-crossover
and related complexes. However, this does not give us a detailed picture
of the host-guest interactions which are at play. A deeper analysis
of these interactions is, in fact, possible via DFT calculations,
such as those which have recently been used to study {[}Fe(bpy)$_{3}${]}$^{2+}$
included in a zeolite Y via the supermolecular model of {[}Fe(bpy)$_{3}${]}$^{2+}$@Y
shown in Fig.~\ref{fig:LSinY} \citep{VHL09}.

Zeolites provide a well-defined rigid three-dimensional network with
a diversity of cavities of different sizes and shapes. Thus the environment
of the complex may be controlled by choosing the cavities in which
the complex is included. The zeolite Y supercage has a diameter of
about 13~Å and an opening of $\approx$7.4~Å which allows the encapsulation
of complexes of the same diameter \emph{via} \emph{in situ} synthesis.
This is the case of \emph{tris}(2,2'-bipy\-ridine) complexes. For
{[}Fe(bpy)$_{3}${]}$^{2+}$@Y, the formation of the LS complex may
be characterized by X-ray diffraction combined possibly with infrared
spectroscopy, UV-visible reflection spectroscopy or solid state NMR
\citep{QPD+82,UMT93,UMT99,VK08}.

Mössbauer absorption spectroscopy of $^{57}$Fe may also be used \citep{QPD+82,VHN+96}.
For iron spin crossover compounds, Mössbauer spectroscopy provides
information about oxidation states, coordination of the central iron
atom, and the distortion of the complexes from octahedral symmetry.
The $C_{3}$ symmetry model shown in Fig.~\ref{fig:LSinY} has been
proposed on the basis of experimental data. DFT calculations have
been carried out for this 229 atom structure for both the HS and LS
states \citep{VHL09}. In particular, the molecular structures of
{[}Fe(bpy)$_{3}${]}$^{2+}$ and of {[}Fe(bpy)$_{3}${]}$^{2+}$@Y
were optimized in their two spin states using various GGA and hybrid
functionals. The optimized HS and LS structures obtained using different
density functionals are very similar. A careful look shows that caging
in the zeolite results in a slight shrinking and a slight distortion
of the {[}Fe(bpy)$_{3}${]}$^{2+}$ structure (Fig.~\ref{fig:LSinY}).
Caging has more effect on the HS state than on the LS state because
the LS state is spatially more compact than is the HS state. The $^{57}$Fe
quadrupolar splitting $\Delta E_{Q}$ provides one measure of the
distortion of the complex. $\Delta E_{Q}$ has thus been calculated
for the {[}Fe(bpy)$_{3}${]}$^{2+}$ complex in isolation and caged
within the zeolite, and it has been found to be correlated with the
predicted degree of distortion of the HS and LS complexes as measured
by the ratio $\xi$ of Fe-N distances (see Ref.~\citep{VHL09}).\footnote{In the gas phase, the complex has the symmetry $D_{3}$ and all the
Fe-N distances are identical: $\xi=1$. {[}Fe(bpy)$_{3}${]}$^{2+}$@Y
has $C_{3}$ symmetry with two different Fe-N distances whose ratio
$\xi<1$ depends upon the degree of distortion of the complex.} Moreover the $\Delta E_{Q}$ values calculated for for LS {[}Fe(bpy)$_{3}${]}$^{2+}$@Y
are in good (even very good) agreement with the experimental value
$\Delta E_{Q}\approx0.32\mbox{ mm.s\ensuremath{^{-1}}}$ of Vankó
\emph{et al.}\citep{VHN+96} % \CITE{ChemCommun_1996_785}
 This emphasizes the pertinence of the model for {[}Fe(bpy)$_{3}${]}$^{2+}$@Y.

\begin{table}
\caption{Calculated HS-LS energy differences $\Delta E_{\mbox{HL}}$ (cm$^{-1}$)
in {[}Fe(bpy)$_{3}${]}$^{2+}$ and in {[}Fe(bpy)$_{3}${]}$^{2+}$@Y,
variation $\Delta\left(\Delta E_{\mbox{HL}}\right)$ of $\Delta E_{\mbox{HL}}$
after being caged in the zeolite, and decomposition of $\Delta(\Delta E_{\mbox{HL}})$
into its geometric part $\Delta E_{\mbox{HL}}^{\mbox{dist}}$ and
the host-guest interaction energy $\Delta E_{\mbox{HL}}^{\mbox{int}}$
The HS state considered here is the trigonal $^{5}$A state of the
ligand field $^{5}T_{2g}$ state. Nevertheless both trigonal $^{5}E$
and $^{5}A$ states are close in energy. Adapted from Ref.~\citep{VHL09}
\label{tab:zeolite:DeltaEHB}}

\centering{}%
\begin{tabular}{lccccc}
\hline 
$\mbox{ }$ & PBE & B3LYP$^{\star}$ & HCTH & O3LYP & OLYP\tabularnewline
\hline 
\multicolumn{6}{c}{HS-LS energy difference $\Delta E_{\mbox{HL}}$}\tabularnewline
$\left[\mbox{Fe(bpy)\ensuremath{_{3}}}\right]^{2+}$ & +10087 & +3849 & +141 & $-811$ & +3660\tabularnewline
$\left[\mbox{Fe(bpy)\ensuremath{_{3}}}\right]^{2+}\mbox{@Y}$ & +12004 & +5925 & +2687 & +1941 & +6594\tabularnewline
\multicolumn{6}{c}{$\Delta(\Delta E_{\mbox{HL}})=\Delta E_{\mbox{HL}}\left[\mbox{Y}\right]-\Delta E_{\mbox{HL}}\left[\varnothing\right]$}\tabularnewline
 & +1917 & +2076 & +2546 & +2752 & +2934\tabularnewline
\multicolumn{6}{c}{$\Delta\left(\Delta E_{\mbox{HL}}\right)=\Delta E_{\mbox{HL}}^{\mbox{int}}+\Delta E_{\mbox{HL}}^{\mbox{dist}}$}\tabularnewline
$\Delta E_{\mbox{HL}}^{\mbox{int}}$ & +1246 & +1267 & +1621 & +1286 & +1331\tabularnewline
$\Delta E_{\mbox{HL}}^{\mbox{dist}}$ & +671 & +809 & +925 & +1466 & +1593\tabularnewline
\hline 
\end{tabular}
\end{table}

DFT provides a very satisfying description of {[}Fe(bpy)$_{3}${]}$^{2+}$@Y
structures and Mössbauer properties in the HS and LS states. However,
as may be anticipated (Sec.~\ref{sec:applications:dft}), and as
shown in the data in Table~\ref{tab:zeolite:DeltaEHB}, evaluating
the HS and LS energies of {[}Fe(bpy)$_{3}${]}$^{2+}$@Y remains a
difficult problem for DFT. In fact, the calculated energy difference
$\Delta E_{\mbox{HL}}[\mbox{Y}]$ varies between +1941 and +12004
cm$^{-1}$ depending upon the functional used. Nevertheless all the
functionals predict that caging destabilizes the HS state more than
the LS state. Furthermore the values found for the variation $\Delta\left(\Delta E_{\mbox{HL}}\right)$
of $\Delta E_{\mbox{HL}}$ are very consistent with each other, all
being between +1917 and +2934 cm$^{-1}$, which is within our expected
accuracy limitations for transition metal complexes. As we shall see
below, the functionals used here permit a quantitative description
of caging effects. The best estimation of $\Delta\left(\Delta E_{\mbox{HL}}\right)$
is thus \citep{VHL09}, 
\begin{eqnarray}
\Delta\left(\Delta E_{\mbox{HL}}\right) & = & \Delta E_{\mbox{HL}}[\mbox{Y}]-\Delta E_{\mbox{HL}}[\varnothing]\nonumber \\
 & = & +2500\pm1000\;\mbox{cm\ensuremath{^{-1}}}\,.\label{eq:zeol:destabilisation}
\end{eqnarray}
The notation $X[\varnothing]$ and $X[\mbox{Y}]$ stand for the values
of $X$ evaluated for the isolated {[}Fe(bpy)$_{3}${]}$^{2+}$ and
for the caged {[}Fe(bpy)$_{3}${]}$^{2+}$@Y respectively. In order
to analyse the influence of host-guest interactions on spin-state
energies, $\Delta\left(\Delta E_{\mbox{HL}}\right)$ has been decomposed
into two contributions: $\Delta E_{\mbox{HL}}^{\mbox{dist}}$ and
$\Delta E_{\mbox{HL}}^{\mbox{int}}$. The first is the HS-LS energy
difference: $\Delta E_{\mbox{HL}}^{\mbox{dist}}=E_{\mbox{HS}}^{\mbox{dist}}-E_{\mbox{LS}}^{\mbox{dist}}$,
where $E_{\Gamma}^{\mbox{dist}}$ ($\Gamma$ = HS, LS) is the energy
needed to deform in the spin state $\Gamma$ the structures of the
complex and the surrounding cage network from their dissociated and
relaxed structures to those found in {[}Fe(bpy)$_{3}${]}$^{2+}$@Y.
The second is the variation of guest-host interaction energy $E^{\mbox{int}}$
induced by the LS $\rightarrow$ HS change of state: $\Delta E_{\mbox{HL}}^{\mbox{int}}=E_{\mbox{HS}}^{\mbox{int}}-E_{\mbox{LS}}^{\mbox{int}}$.

Table~\ref{tab:zeolite:DeltaEHB} shows that the value of $\Delta E_{\mbox{HL}}^{\mbox{int}}$
is relatively independent of the functional used ($\Delta E_{\mbox{HL}}^{\mbox{int}}\approx\mbox{+1300 cm\ensuremath{^{-1}}}$,
contrary to the case for $\Delta E_{\mbox{HL}}^{\mbox{dist}}$. The
variation in the values of $\Delta E_{\mbox{HL}}^{\mbox{dist}}$ is
thus responsible for the spread of values in $\Delta\left(\Delta E_{\mbox{HL}}\right)$.
This can be traced back to the influence of the functional on the
description of the complex caged in the zeolite~Y supercage \citep{VHL09}.
$\Delta E_{\mbox{HL}}^{\mbox{dist}}>0$ because the distortional energy
is greater in the HS than in the LS state. Also $\Delta E_{\mbox{HL}}^{\mbox{int}}>0$
as the host-guest interactions become less stabilizing after the LS
$\rightarrow$ HS transition. The instantaneous interaction energy
$E^{\mbox{int}}$ between the two fragments may be further decomposed
\citep{M77,VBB+01} as, 
\begin{equation}
E^{\mbox{int}}=E^{\mbox{elstat}}+E^{\mbox{Pauli}}+E^{\mbox{orb}}\,.\label{eq:interaction-energy:contributions}
\end{equation}
$E^{\mbox{elstat}}$ is the classical electrostatic energy between
the unperturbed charge distributions of the host and guest; $E^{\mbox{Pauli}}$
is the Pauli repulsion responsible for steric repulsions; and $E^{\mbox{orb}}$
is the orbital energy contribution, i.e., the energy obtained upon
relaxation of the electron density. More precisely, the Pauli repulsion
is the change in the energy in going from a simple superposition of
the unrelaxed density of the two fragments to the wavefunction of
the two fragments together obtained by antisymmetrization and normalization
of the product of the fragment wavefunctions. Table~\ref{tab:zeolite:interactions}
shows the results obtained by this analysis of the host-guest interactions
in {[}Fe(bpy)$_{3}${]}$^{2+}$@Y. Notice that the interactions are
binding in the two spin states ($E_{\Gamma}^{\mbox{int}}<0$) and
that the binding is more electrostatic than covalent ($E_{\Gamma}^{\mbox{elstat}}<E_{\Gamma}^{\mbox{orb}}<0$;
$\Gamma=$ HS, LS). In going from the LS to the HS state, the change
in the interaction energy is $\Delta E_{\mbox{HL}}^{\mbox{int}}$
= +1248 cm$^{-1}$, in very good agreement with the value given in
Tab.~\ref{tab:zeolite:DeltaEHB}. The largest contribution to $\Delta E_{\mbox{HL}}^{\mbox{int}}$
comes from the Pauli repulsion, which is intuitively consistent with
the increase in the size of the complex in going from the LS to the
HS state. However $\Delta E_{\mbox{HL}}^{\mbox{Pauli}}$ only represents
about 53\% of $\Delta E_{\mbox{HL}}^{\mbox{int}}$: The electrostatic
and orbital interaction energies are less stabilizing and contribute
to 47\% of $\Delta E_{\mbox{HL}}^{\mbox{int}}$. This is more difficult
to rationalize based upon chemical intuition alone.

\begin{table}
\caption{Analysis of the host-guest interactions in {[}Fe(bpy)$_{3}${]}$^{2+}$
in the HS and LS states (energies in cm$^{-1}$ for the OLYP functional)
\citep{VHL09} \label{tab:zeolite:interactions}}

\centering{}%
\begin{tabular}{ccccc}
\hline 
 & $E_{\Gamma}^{\mbox{elstat}}$ & $E_{\Gamma}^{\mbox{Pauli}}$ & $E_{\Gamma}^{\mbox{orb}}$ & $E_{\Gamma}^{\mbox{int}}$\tabularnewline
\hline 
$\Gamma$ = LS & $-15558$ & +17442 & $-12158$ & $-10274$\tabularnewline
$\Gamma$ = HS & $-15251$ & +18099 & $-11874$ & $-9026$\tabularnewline
\hline 
\hline 
 & $\Delta E_{\Gamma}^{\mbox{elstat}}$ & $\Delta E_{\Gamma}^{\mbox{Pauli}}$ & $\Delta E_{\Gamma}^{\mbox{orb}}$ & $\Delta E_{\Gamma}^{\mbox{int}}$\tabularnewline
\hline 
 & +307 & +657 & +284 & +1248\tabularnewline
\hline 
\end{tabular}
\end{table}

In fact, for a given geometry {[}M(bpy)$_{3}${]}$^{2+}$@Y where
M$^{2+}$ is a doubly-charged transition metal cation, the host-guest
interaction energies are independent of the metal cation spin state.
It follows that the host-guest interactions in the {[}M(bpy)$_{3}${]}$^{2+}$@Y
inclusion compound are \emph{closed-shell interactions between the
first coordination sphere of the complex and the second coordination
sphere constituted by atoms of the surrounding zeolite, polarized
by the doubly-charged cation} \citep{VHL09}. The good performance
of most modern functionals for the description of interactions between
closed shells explains the good performance obtained. As concerns
spin-state energies, the best estimate of $\Delta E_{\mbox{HL}}[\varnothing]$
obtained from CASPT2 calculations is 3700 $\pm$ 1000 cm$^{-1}$ \citep{PV06,PV08}.
Combining this result and that of Eq.~(\ref{eq:zeol:destabilisation})
gives 
\[
\Delta E_{\mbox{HL}}[\mbox{Y}]=6200\pm1500\mbox{ cm\ensuremath{^{-1}}}\,,
\]
as the best estimate of the energy difference between the two spin
states in {[}Fe(bpy)$_{3}${]}$^{2+}$@Y and a difference of zero
point energies of $\Delta E_{\mbox{HL}}^{\circ}[\mbox{Y}]\approx5400\mbox{ cm\ensuremath{^{-1}}}$
\citep{VHL09}. In the theory of HS $\rightarrow$ LS relaxation presented
above \citep{HEL+06}, the value of $\Delta E_{\mbox{HL}}^{\circ}[\mbox{Y}]$
corresponds to a lifetime $\tau_{\mbox{HS}}=1/k_{\mbox{HL}}(T\rightarrow0\,\mathrm{K})$
of the HS state of 10~ns at most. This agrees with the limit $\tau_{\mbox{HS}}[\mbox{Y}]\leq60\mbox{ ns}$
deduced from Mössbauer emission spectroscopy \citep{HyperfineInteract_126_163}.

The study of {[}Fe(bpy)$_{3}${]}$^{2+}$@Y shows that DFT allows
environmental effects to be taken into account in an efficient and
reliable fashion. The DFT calculations have indeed allowed the quantification
of the effect of caging on the structure of the complex in the HS
and LS states, the determination of the relative energies of the spin
states, and finally a detailed analysis of guest-host interactions.
Furthermore periodic DFT studies offer interesting possibilities for
a deeper understanding of the effects of crystal packing and of counter
ions in the solid state. Such studies can also help gain insights
into how sorption of guest molecules modulate the spin state of spin-crossover
complexes encaged in porous metal-organic frameworks, a mechanism
developed for the design of a novel spin-crossover porous hybrid architecture
for potential sensing applications \citep{ChemCommun_55_194}.

\begin{figure}[!tbh]
\centering{}\hfill{}%
\begin{minipage}[c][1\totalheight][t]{0.5\columnwidth}%
\begin{center}
\includegraphics[scale=0.3]{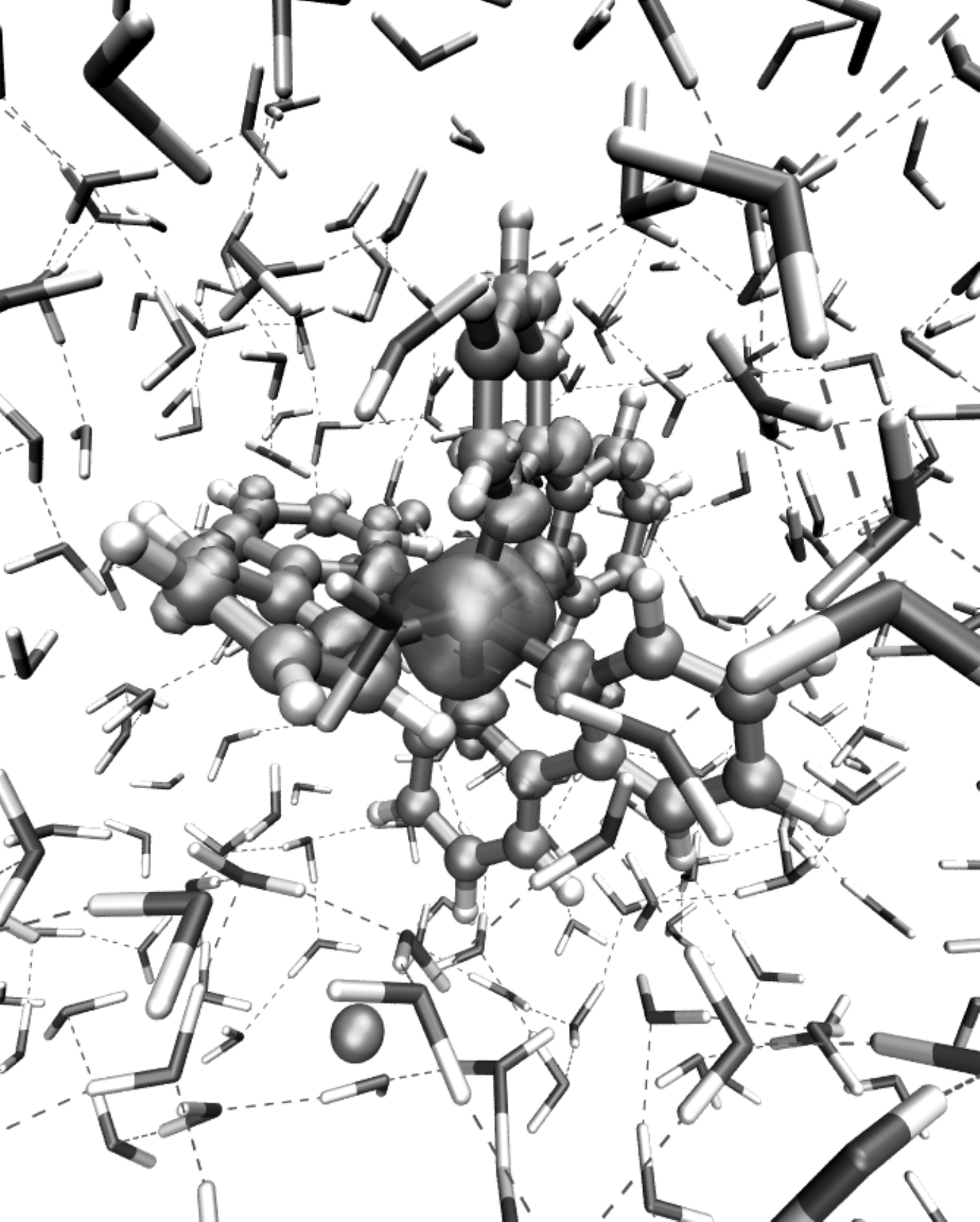}
\par\end{center}%
\end{minipage}\hfill{}%
\begin{minipage}[c]{0.35\columnwidth}%
\begin{center}
\includegraphics[scale=0.35]{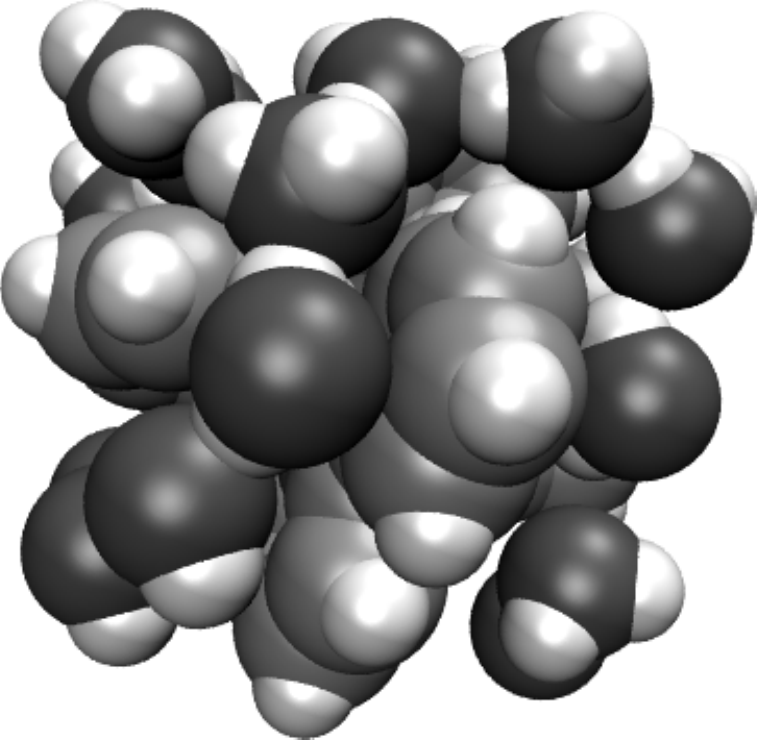}
\par\end{center}%
\end{minipage}\hfill{}\caption{Car-Parrinello molecular dynamics study of {[}Fe(bpy)$_{3}${]}(Cl)$_{2}$
at 300\,K in the LS and HS states Ref.~\citep{LH10}. Left: snapshot
from the HS {[}Fe(bpy)$_{3}${]}$^{2+}$ trajectory showing {[}Fe(bpy)$_{3}${]}$^{2+}$,
a Cl$^{-}$ anion, water molecules, and the spin-density on the iron.
Right: snapshot from the LS trajectory showing that the first solvation
shell around the complex is made up of water molecules inserted into
the spaces between the ligands. Figure adapted from Ref.~\citep{LH10}.
\label{fig:febpy:solution-aq}}
\end{figure}

Similarly the study of Fe(II) complexes in solution by molecular dynamics
simulations allowed to progress in the comprehension of the spin-state
dependence of their structural and vibrational properties, evidencing
the strong interpenetration between their coordination sphere and
their hydration shell \citep{LH10,PhysChemChemPhys_20_6236,PhysChemChemPhys_21_650}.
This study has been initiated to complement experiment. The mechanism
of the photoinduced LS $\rightarrow$ HS change of states is indeed
extensively investigated for spin-crossover and LS Fe(II) complexes
in solution using ultrafast X-ray, \citep{JPhysChemA_110_38,PhysRevLett_98_057401,JChemPhys_130_124520,Science_323_489,AngewChemIntEd_49_5910,JAmChemSoc_132_61,JAmChemSoc_132_6809,JPhysChemLett_2_880,JPhysChemA_116_9878,JPhysChemA_117_735,JPhysChemC_118_4536,Nature_509_345,JPhysChemC_119_3312,JPhysChemC_119_3322,JPhysChemC_119_5888,JPhysChemB_120_1158,JPhysChemLett_7_465,NatCommun_8_15342,JAmChemSoc_139_17518}
UV-vis \citep{JAmChemSoc_122_4092,JAmChemSoc_129_8199,JAmChemSoc_130_14105,InorgChimActa_361_3937,AngewChemIntEd_48_7184,NatureChem_7_629,ChemPhysChem_18_465}
and vibrational \citep{InorgChem_43_4289,PhysChemChemPhys_10_4264,JAmChemSoc_130_14105}
spectroscopies. Much attention is paid to the prototypical {[}Fe(bpy)$_{3}${]}$^{2+}$
complex in water (Fig.~\ref{fig:febpy:solution-aq}), which has been
the subject of several molecular dynamics studies \citep{LH10,JChemPhys_136_064519,JPhysChemB_120_206,PhysChemChemPhys_18_4789,PhysChemChemPhys_21_4082}.
Combining forefront spectroscopy studies and state-of-the-art molecular
dynamics simulations really proved to be a promising approach for
gaining insights into the mechanism of the photoinduced LS $\rightarrow$
HS change of states. This recently allowed visualizing the coordination
spheres of photo-excited aqueous {[}Fe(bpy)$_{3}${]}$^{2+}$ \citep{PhysChemChemPhys_21_9277}.

\section{The Future\label{sec:the-future}}

The last two decades have seen major progresses in the modelling of
spin-crossover complexes. This may be better apprehended by recalling
the three Pauling points presented in the beginning of the chapter.
Thus, it must first be emphasized that LFT (the first point) is still
very important in the sense that it is inextricably linked to the
way experimentalists think about transition metal complexes. Advances
beyond LFT have come from the concerted effort of several groups which
have converged to establish the second Pauling point (DFT and TD-DFT.)
Not only does DFT perform well for the calculation of geometries,
vibrational frequencies, and other properties of LS and HS complexes,
but DFT is increasingly able to give information about the relative
energies of the different spin states. This is partly due to the development
of specific procedures, including the reparameterization of hybrid
functionals, the discovery of useful functionals on the lower rungs
of Jacob's ladder, and the discovery of functional-independent invariants
which can allow calculations to be less dependent on the choice of
functional. Progress also concerns the third Pauling point and a synergy
has begun to emerge between the second and third Pauling points. Thus
coupled-cluster calculations are now able to validate DFT results
for the relative energies of different spin states for small- to medium-sized
complexes. In the future, the authors expect to see increasing number
of coupled-cluster and large-active-space CASSCF/CASPT2 calculations
on spin-crossover complexes. Nevertheless DFT and TD-DFT will continue
to be their main tools, particularly for the study of large systems
and especially when it is necessary to take into account environmental
effects and/or photodynamics. This chapter has presented some of the
work of the authors going in this direction.

\bibliographystyle{elsarticle-num}
\bibliography{chapter_MEC-LDLM}

\end{document}